\documentclass[sigconf,twocolumn, 9pt]{acmart}

\graphicspath{{./pix/}}
\DeclareGraphicsExtensions{.pdf, .eps, .mps, .png, .jpg}
\usepackage[english]{babel}
\usepackage{adjustbox}
\usepackage{tabularx,enumitem,microtype}
\usepackage{booktabs,amsfonts}
\usepackage[ruled,vlined,linesnumbered]{algorithm2e}
\usepackage{endnotes,hyperref,color}
\usepackage[noend]{algpseudocode}
\usepackage{gensymb,xspace,hyphenat}
\usepackage{listings,xcolor,graphicx}
\usepackage[labelformat=simple]{subcaption}
\usepackage{tikz}
\usepackage{tcolorbox}
\usepackage{pifont}
\usepackage{times}
\usepackage{balance}
\usepackage[justification=centering]{caption}

\def\circled#1{{\ooalign{\hfil\lower.1ex\hbox{#1}\hfil\crcr\Orb}}}

\def\HiLi{\leavevmode\rlap{\hbox to \linewidth{\color{black!15}\leaders\hrule height .8\baselineskip depth .5ex\hfill}}}

\makeatletter
\newcommand{\algcolor}[2]{%
  \hskip-\ALG@thistlm\colorbox{#1}{\parbox{\dimexpr\linewidth-2\fboxsep}{\hskip\ALG@thistlm\hspace*{-3pt}\vspace*{-3pt}#2}}%
}

\newcommand{\change}[1]{{\color{black}{#1}}}
\makeatother

\newcommand{\systemname}{REITS\xspace}

\newcommand{\ie}{{\it i.e.,}\xspace}

\newcommand{\parahead}[1]{\vspace*{4.5pt plus 2pt minus 2pt}\noindent %
  {\bfseries #1}}

\renewcommand{\paragraph}[1]{\parahead{#1}}

\newcounter{packednmbr}

\setcopyright{acmcopyright}
\copyrightyear{2021}
\acmYear{2021}

\setitemize{itemsep=1pt,topsep=2pt,parsep=1pt,partopsep=0pt,leftmargin=1em}
\setenumerate{itemsep=1pt,topsep=2pt,parsep=1pt,partopsep=0pt,leftmargin=1.5em}
\setlist{itemsep=1pt,parsep=1pt}

\acmConference[HotMobile 2021]{The 22nd International Workshop on Mobile Computing Systems and Applications}{February 24--26, 2021}{Virtual, United Kingdom}
\acmBooktitle{The 22nd International Workshop on Mobile Computing Systems and Applications (HotMobile 2021), February 24--26, 2021, Virtual, United Kingdom}
\acmPrice{15.00}
\acmDOI{10.1145/3446382.3448650}
\acmISBN{978-1-4503-8323-3/21/02}

\begin{document}



\title{\systemname: Reflective Surface for Intelligent Transportation Systems} 

\author{Zhuqi Li$^*$}
\affiliation{%
  \institution{Princeton University}
}
\email{zhuqil@cs.princeton.edu}

\author{Can Wu$^*$}
\affiliation{%
  \institution{Princeton University}
}
\email{canw@princeton.edu}

\author{Sigurd Wagner}
\affiliation{%
  \institution{Princeton University}
}
\email{wagner@princeton.edu}

\author{James C. Sturm}
\affiliation{%
  \institution{Princeton University}
}
\email{sturm@princeton.edu}

\author{Naveen Verma}
\affiliation{%
  \institution{Princeton University}
}
\email{nverma@princeton.edu}

\author{Kyle Jamieson}
\affiliation{%
  \institution{Princeton University}
}
\email{kylej@cs.princeton.edu}





\begin{abstract}
Autonomous vehicles are predicted to dominate the transportation industry in the foreseeable future. Safety is one of the major challenges to the early deployment of self-driving systems. To ensure safety, self-driving vehicles must sense and detect humans, other vehicles, and road infrastructure accurately, robustly, and timely. However, existing sensing techniques used by self-driving vehicles may not be absolutely reliable. 
In this paper, we design \systemname, a system to improve the reliability of RF-based sensing modules for autonomous vehicles. 
We conduct theoretical analysis on possible failures of existing RF-based sensing systems.
Based on the analysis, \systemname adopts a multi-antenna design, which enables constructive blind beamforming to return an enhanced radar signal in the incident direction. \systemname can also let the existing radar system sense identification information by switching between constructive  beamforming state and  destructive  beamforming state. Preliminary results show that \systemname improves the detection distance of a self-driving car radar by a factor of 3.63.
{\let\thefootnote\relax\footnote{{$*$ Z.L. and C.W contributed equally to this paper}}}
\end{abstract}
    %

\ccsdesc[500]{Networks~Mobile networks}
\ccsdesc[300]{Networks~Mobile and wireless security}
\ccsdesc[500]{Hardware~Hardware reliability}

\keywords{Self-driving car, Robustness, Surface}

\maketitle

\section{Introduction}
Autonomous vehicles are becoming ever more attractive due to their potential to enable more efficient transportation by reducing traffic congestion, fuel consumption, and air pollution. To realize  widespread adoption, autonomous vehicles must meet extremely high safety standards. For example, an autonomous driving system that is comparable to human drivers should achieve a failure rate of less than 1.09 fatalities per 100 million miles~\cite{kalra2016driving}.

\begin{figure}[t]
\includegraphics[width=1\linewidth]{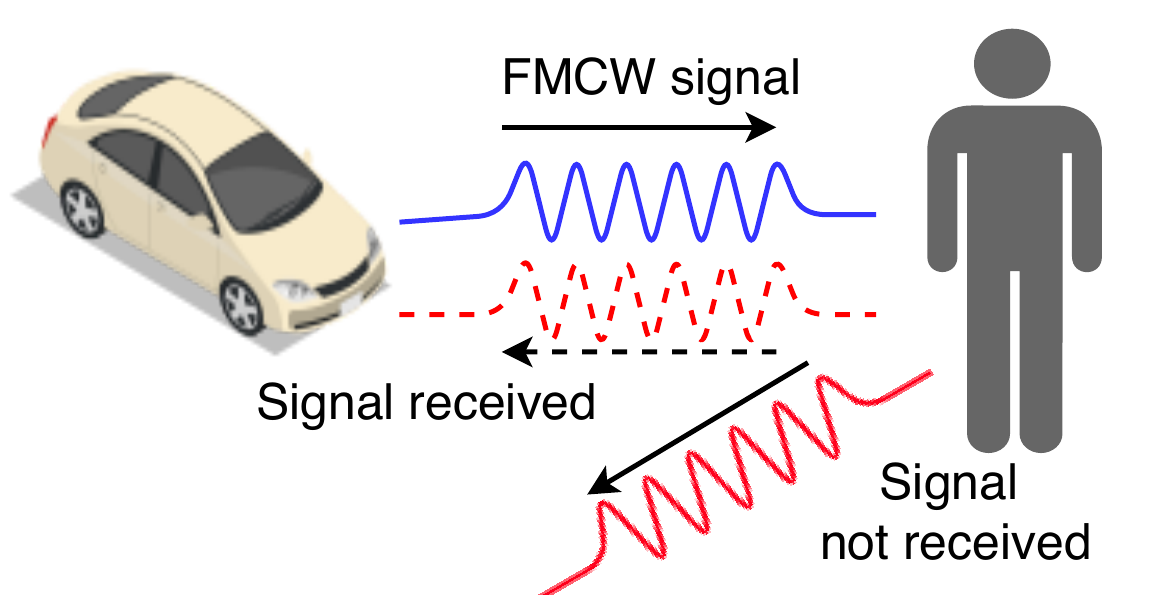}
\caption{Self-driving car radar detects objects of  interest by measuring the reflected signal. But the majority of radar signal might not be reflected back to the incident direction.}
\label{fig:radar_example}
\end{figure}

In order to meet the high safety standard, a lot of previous works have improved the autonomous control
systems with deep learning based algorithms~\cite{kuderer2015learning}. However, the reliability of those algorithms heavily depends on the signal input from sensors. A safe autonomous vehicle must reliably sense humans, other vehicles, and proximate road infrastructure accurately and robustly. Therefore, autonomous vehicles are typically equipped with two categories of sensors. One category is light-based sensors, \ie cameras or Lidar. However, light-based sensors suffer from excessive absorption or scattering of the probing signal during inclement weather. Specifically, light-based sensors may fail when light is blocked or scattered by fog, rain, or snow. The other category is RF-based sensors, \ie radar. RF-based sensors are more reliable in extreme weather, but they also may fail to detect the object of interest when the surface of the target is not perpendicular to the incident wave. In such a case, most of the probing signal is reflected to a direction different from the incident wave, making the radar system receive a subthreshold return signal. This causes irrelevant reflections from the environment to dominate the return signal and significantly degrade its SNR. Since both types of sensors are not absolutely reliable, the self-driving system, therefore, may fail to sense the surrounding environment, which often leads to deadly consequences~\cite{tesladeath}. \change{ The signal reflection pattern only depends on the shape and orientation of the object, which means optimization at the radar is not going to solve the problem. The natural solution is to change the radar reflection pattern of the target object's surface. 
}

\change{
The main challenge to achieve this goal is the dilemma between signal strength requirement and the low power constraint. On the one hand, the solution must deliver significant energy to the direction of radar receiver so that radar system can get signal with significant high SNR. On the other hand, the solution must be low power since it has to attach to the surface of any object, which excludes the design choices of using active electronic elements. Another challenge for self-driving car radar is the limitation on its sensing capability. Although self-driving radar can sense the distance, angle, and velocity of the target object, it typically cannot get the identification information (e.g. whether a traffic sign is a stop sign or speed limit sign). The limitation on radar sensing capability further reduces the robustness of self-driving system in extreme weather when vision based sensing does not work. 
}

\change{To solve those challenges,  we propose \systemname, a system designed for robust detection of objects by RF-based sensing. }
\systemname augments existing radar systems for self-driving vehicles, and includes two design components: (1) a surface based on a Van Atta type array reflector~\cite{sharp1960van} that beamforms the return signal from the object by constructive interference, and (2) programmable RF switches that alternate between constructive and destructive beamforming. When set to constructive interference, surface beamforms the signal back to its incident direction so that the self-driving radar gets high SNR. 
When set to destructive interference, signal absorbed by different parts of the surface cancel out to each other, leaving no reflection to the direction of incoming signal.
By switching between constructive beamforming state and destructive beamforming state, \systemname can encode object identification information so that the self-driving car radar can get more detail information about the target (e.g. whether the object is a vehicle or pedestrian).
Wide-spread deployment of \systemname in self-driving vehicle applications is anticipated from its reduced complexity and cost, because only passive antennas and switches are needed in the design, in contrast to conventional radar reflectors based on sophisticated active electronics \cite{ActiveReflector,ActiveBackscatterTag}.

By attaching \systemname to the surface of vehicles, road infrastructures, and clothes, commercial self-driving systems will be able to detect objects of interest in a more reliable manner. To achieve such a design, we make the following key contributions in this paper:

\begin{itemize}
    \item We study the root cause of RF-based sensor failure for self-driving vehicles and propose an RF circuit design based on a Van Atta type array, which automatically beamforms the signal back to the radar receiver of self-driving vehicles.
    \item We propose a method to transmit object identification information from the surface to self-driving car radar
    by switching between constructive beamforming state and destructive beamforming state.
    \item We verify the design of \systemname with SIMULIA CST, a time-domain 3D full-wave electromagnetic solver based on finite integration and conduct a preliminary evaluation of the performance of \systemname on 24-Ghz self-driving radar band. The results show that a four-antenna \systemname improves the signal strength by 11.2~dB and correspondingly extends the radar detection range by $3.63\times$.
\end{itemize}



\section{Failure Analysis for Self-driving Radar}
\label{ss:failurecause}
Self-driving radar uses Frequency-Modulated Continuous-Wave Radar (\textit{FMCW}) technique to detect objects of interest. As shown in Figure~\ref{fig:radar_example}, the radar sends a chirp signal that swipes a wide frequency band as the probing signal. After sending the chirp signal, the radar captures the signal reflected by the objects of interest. If the reflected signal strength exceeds a threshold, the self-driving radar can calculate the distance, direction, and speed of the target objects by applying radar detection algorithm. Since the radar signal receiver and transmitter are co-located at the same vehicle, the radar system can only capture the signal that is reflected back in the incident direction. As shown in Figure~\ref{fig:radar_example}, the signal might also be reflected into a different direction, which cannot be received by the radar receiver. In order to have a reliable sensing capability, the signal captured by the radar receiver must always exceed the detection threshold. 

\begin{figure}[t]
\includegraphics[width=0.9\linewidth]{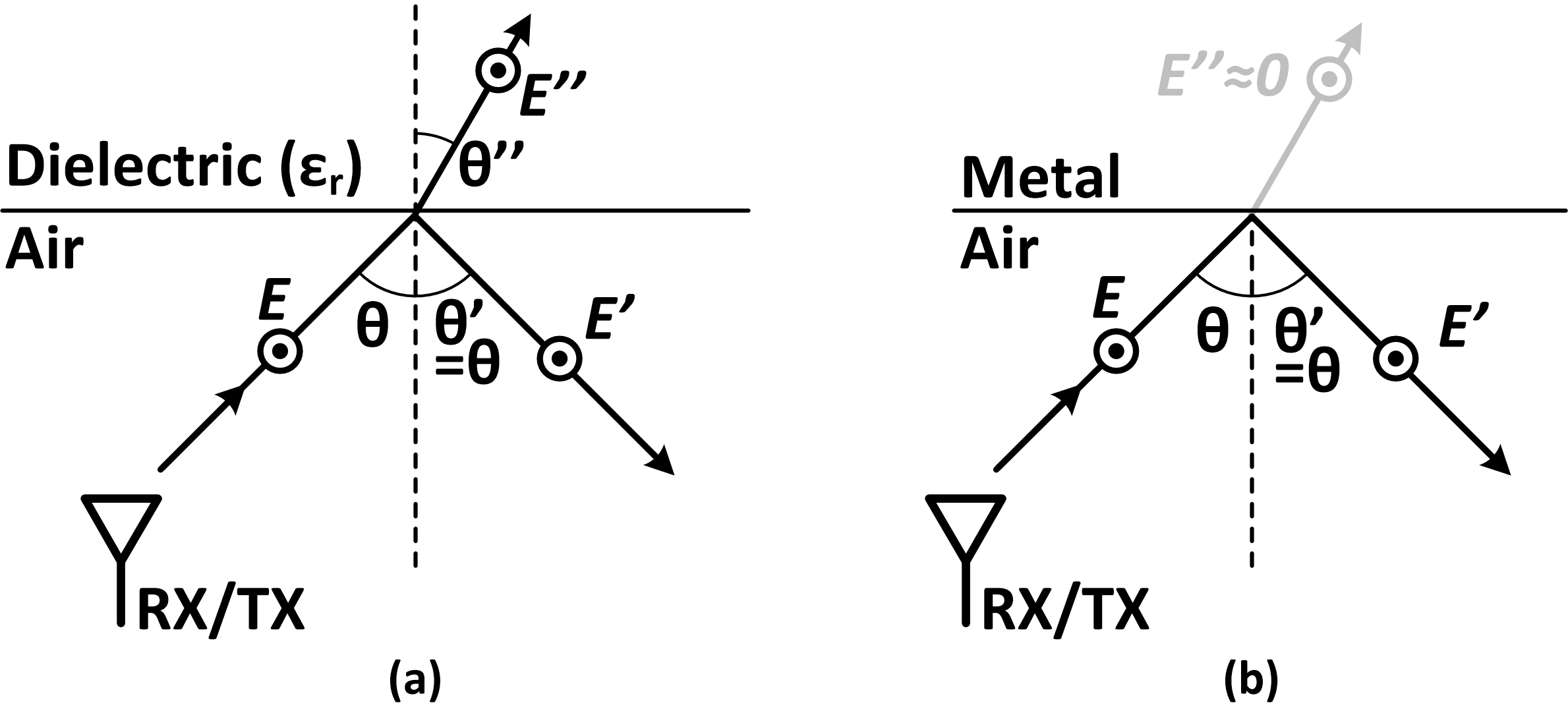}
\caption{Reflection of an RF signal at \\ objects made of (a) dielectric; (b) metal.}
\label{fig:snell_law}
\end{figure}

The signal reflection follows Snell's Law, as shown in Figure~\ref{fig:snell_law}. If the object is made of a dielectric material, part of the incoming RF signal penetrates and the other part bounces off the surface and heads in a different direction, symmetric by the surface normal to the incoming angle, {\it i.e. ,} $\theta'=\theta$. When the object is made of metal, nearly the entire RF signal ($E'\approx E$) bounces off like a ball hitting a wall. In order to have the energy reflected back to the original direction, the  incoming angle of the signal has to be zero ($\theta=0^o$). But in most cases, the surface of the target is not perpendicular to the incoming signal. In such cases, only a small portion of the RF signal returns along the incoming path, resulting in a small signal at the receiver. The signal strength returning to the radar receiver becomes weak, which significantly degrades the detectable distance of radar systems and leads to failure of radar detection.

\begin{figure*}
      \begin{minipage}[h]{0.27\linewidth}
         \centering
         \includegraphics[width=1\linewidth]{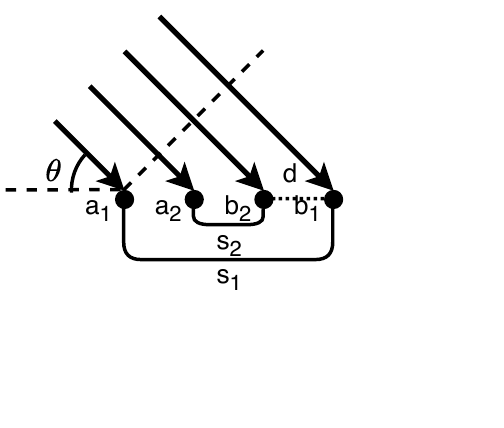}
        \caption{The blind beamformer as a one-dimensional array.}
        \label{fig:antenna_array}
      \end{minipage}
      \hfill
      \begin{minipage}[h]{0.355\linewidth}
         \centering
         \includegraphics[width=1\linewidth]{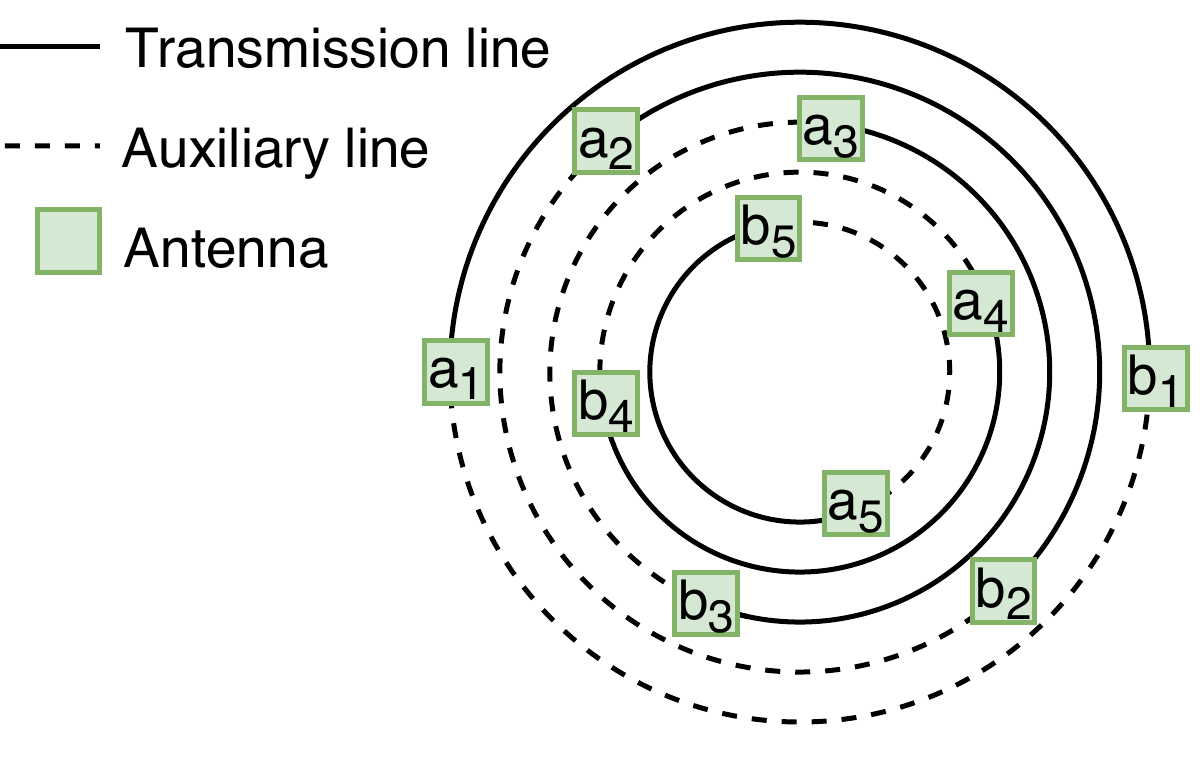}
        \caption{The blind beamformer \\ as a two-dimensional surface.}
        \label{fig:antenna_surface}
      \end{minipage}
      \hfill
      \begin{minipage}[h]{0.355\linewidth}
         \centering
         \includegraphics[width=1\linewidth]{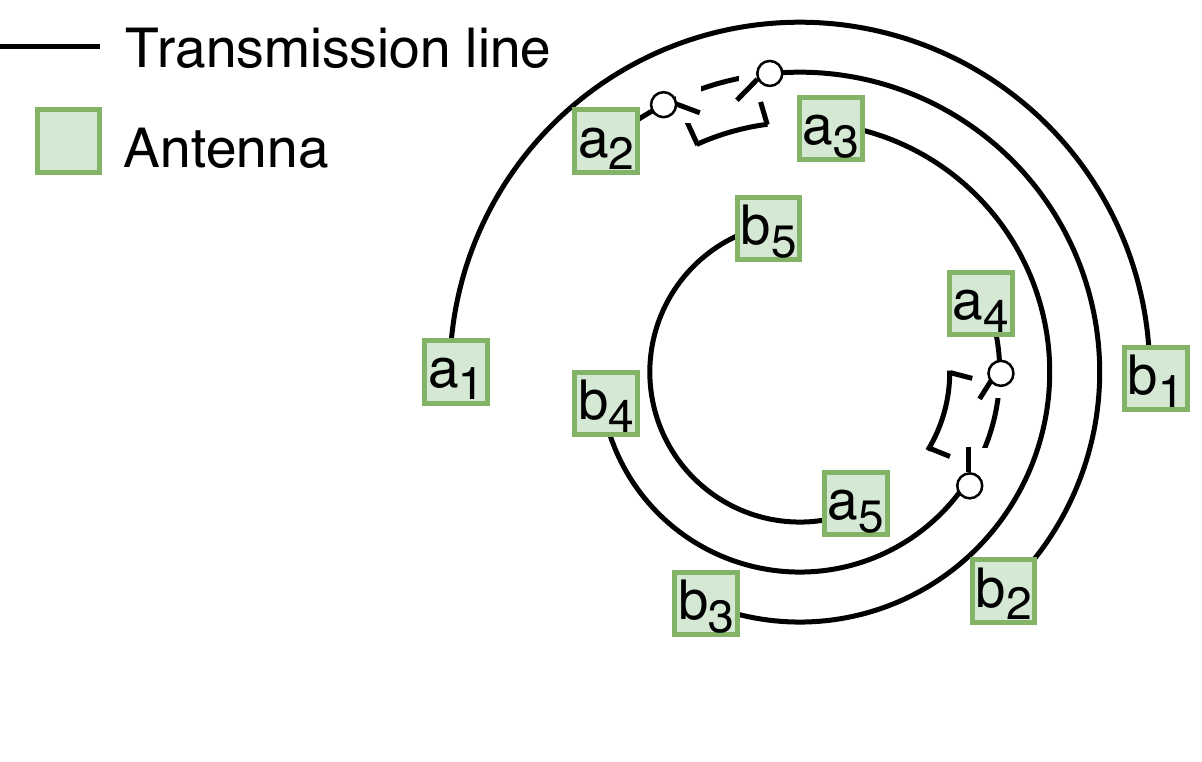}
        \caption{The blind beamformer \\ with programmability.}
        \label{fig:antenna_surface_pro}
      \end{minipage}
\end{figure*}

Based on the analysis, the returning signal strength of the self-driving radar depends on the area of the surface that is perpendicular to the incoming wave. Such impact to the robustness of radar signal detection is typically modeled as Radar Cross Sector (\textit{RCS}). An object may not always have a large RCS in the view of self-driving car radar. In a real road scenario, the RCS of a target depends on its position, shape, and orientation. Therefore, the RCS of the object changes constantly along with dynamic road environment, leading to uncertainty in radar detection.

\systemname is designed to eliminate the uncertainty of RF-based sensing for self-driving vehicles. It makes the target object have consistently good RCS regardless of its position, shape, and material so that the radar receiver can always capture strong reflection from the object of interest for sensing and detection.




\begin{figure}[t]
\includegraphics[width=1\linewidth]{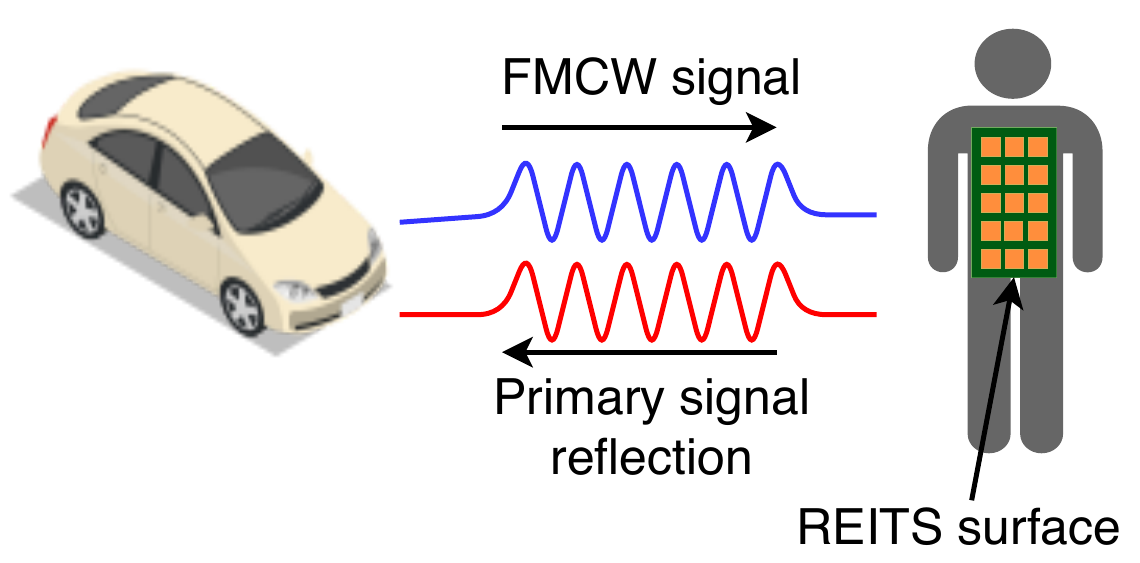}
\caption{\systemname enables robust sensing for self-driving car radar by reflecting the radar signal to its incoming direction.}
\label{fig:radar_example_reits}
\end{figure}

\section{\systemname Design}
\systemname is a programmable surface that can be attached to the surface of an object so that it can have consistently high RCS in the view of the self-driving car radar. As shown in Figure~\ref{fig:radar_example_reits}, \systemname can reflect most of the signal back in its incoming direction so that the radar system can collect a high SNR signal for sensing and detection. Augmented with programmability, \systemname can also transmit object identification information with on-and-off keying modulation.  

To realize such a goal, \systemname' design includes two critical components: a programmable surface that can beamform the radar signal back in the incident direction (\S\ref{ss:surface}), and a mechanism that transmits object identification information to the radar system (\S\ref{ss:algorithm}). \systemname' design does not require any hardware modifications to the existing self-driving radar system. It is anticipated to be deployed where a maximal RCS is desired for enhancing radar detection, such as pedestrians, cyclists, and vehicles.  In addition, the design of \systemname does not involve any  electronic elements with high power consumption (e.g. amplifier or ADC) so that the whole surface can operate at an extreme low-power constraint.





\subsection{Design of the programmable surface}
\label{ss:surface}
With mirror-like reflection (as shown in Figure~\ref{fig:snell_law}), the majority of the TX signal is reflected at an angle different from the incoming direction (unless the incident wave is perpendicular to the target surface). In order to maximize the radar cross sector of a target surface, we need a mechanism that controls the propagation path of the signal.

An obvious approach is to collect the energy, manipulate the signal, and send it back. However, a single antenna does not collect sufficient power to amplify the RCS for two reasons: (1) the millimeter-wave signal used by self-driving car radars experiences high propagation loss; (2) antennas operated at millimeter-wave band are small in size, typically $1.97$-$6.25$~mm (half wavelength for $24-76$~GHz radio). Such small antennas do not receive enough power for enabling active operations such as signal amplification.  

To raise the intercepted power, the target can be equipped with an array of antennas, where the array beamforms the signal back into the incident direction. However, implementing beamforming by traditional methods requires either multiple radio chains~\cite{shepard2012argos} or a phased array~\cite{mailloux1982phased}, which typically exceeds the energy budget of a low-power device. \systemname takes a different approach by implementing a blind beamforming design to enable an array of antennas beamforming the radar signal back to its incident direction. We first illustrate the idea of blind beamforming with a one-dimensional array (\S\ref{sss:beamform1d}) and then scale up the design to a two-dimensional surface (\S\ref{sss:beamform2d}). We also introduce programmability to the surface, for signal modulation (\S\ref{sss:beamformProgram}).



\subsubsection{Blind beamforming array}
\label{sss:beamform1d} \systemname adopts a blind beamformer that makes the return signals constructively interfere in the incident direction. Figure~\ref{fig:antenna_array} is an example of a blind beamformer design using a linear four-antenna array. $a_1,a_2,b_1,b_2$ are antennas, and $s_1, s_2$ are transmission lines. The gap between each pair of adjacent antennas is $d$. Suppose the distance between signal source  and $a_1$ is $L_{a_1}$. In practice, the incident wave can be approximated as a plane wave since $L_{a_1}\gg d$. The distances between the signal source and $a_2, b_2, b_1$ can be expressed as $L_{a_1}+d sin(\theta), L_{a_1}+2d sin(\theta), L_{a_1}+3d sin(\theta)$ respectively. The total length of signal propagation path that goes through antenna $a_1$ and antenna $b_1$ is 
$$2L_{a_1} + 3d sin(\theta) + l_{s_1},$$ 
where $l_{s_1}$ is the electrical length of the transmission line $s_1$. Similarly, the length of the path that goes through antenna $a_2$ and antenna $b_2$ is 
$$2L_{a_1} + 3d sin(\theta) + l_{s_2}.$$
We can control the transmission line length $l_{s_1}$ and $l_{s_2}$ in the board design so that they differ by a wavelength. With such a design, the signals from the four antennas constructively interfere in the incident direction.

The characteristic of the blind beamformer can be generalized as long as the configuration of the antenna array satisfies the following two requirements: 
\begin{itemize}
  \item All pairs of antennas are centrosymmetric to the same central point. This requirement ensures that the total distance between the signal source and any pair of antennas $(a_i, b_i)$, $L_{a_i}+L_{b_i}$, equals twice the distance between the signal source and the central point. Therefore, this part of path length is independent of the incident angle, making the beamformer work for any incident direction. 
  \item The length difference of any two transmission lines $s_i$ and $s_j$ satisfies $|l_{s_i} - l_{s_j}|=k \lambda$, where $k$ is an integer and  $\lambda$ is the wavelength. This requirement ensures that any two propagation paths have identical phase delays. Therefore, the signals that travel through all propagation paths constructively interfere with each other at the radar receiver. 
\end{itemize}



\subsubsection{Scale to a two-dimensional surface.}
\label{sss:beamform2d}
We proceed to scale up the blind beamforming design from a one-dimensional array to a two-dimensional array. To harvest sufficient energy for radar detection, the deployment of antennas on the surface should be dense. 
Figure~\ref{fig:antenna_surface} shows the design of the two-dimensional surface for \systemname. 
All antennas are placed on concentric circles. Any pair of antennas is connected by a 180-degree arc transmission line placed on one of the concentric circles. The radius difference of any two circles is $\frac{k\lambda}{\pi}$, which ensures that the lengths of the transmission lines differ by $k\lambda$. 
In addition, the antennas are spaced apart by at least $\lambda/2$ to ensure minimal crosstalk. 


\subsubsection{Enable programmability.}
\label{sss:beamformProgram}
\systemname can be made programmable for controlling the reflected signal strength. As shown in Figure~\ref{fig:antenna_surface_pro}, for this purpose, \systemname is provided  with RF switches on the transmission lines that are labeled with even indexes. Switching-in a second propagation path increases the length of the transmission line by half a wavelength. Therefore, the phase of the signal increases by $\pi$. This mechanism enables destructive interference between different antenna pairs. By changing the number of paths that are switched into the destructive phase, the surface controls the total signal strength that is reflected into the incident direction from the target.

\subsection{Transmitting object identification information}
\label{ss:algorithm}
Besides from the location and velocity, self-driving system also requires object identification information to operate safely. For example, it needs to know whether a traffic sign is a stop sign or a speed limit sign. \systemname can help the existing radar system collect such kind of information. 
Specifically, \systemname switches between constructive beamforming state and destructive beamforming state, so that the radar can receive the signals at the two different states. 
In the constructive beamforming state, the radar receives the signal reflection from \systemname. In the destructive beamforming state, \systemname inserts an extra phase of 180 degrees to half of the signals, by reconfiguring the switches that alter the lengths of the transmission lines (as shown in Figure~\ref{fig:antenna_surface}). It forms destructive beamforming in the incident direction. As a result, the radar cannot receive the reflection from \systemname. \systemname can therefore modulate the object identification information into these two states. When turned into constructive beamforming state, the surface transmits bit $1$. When turned into destructive beamforming state, the surface transmits bit $0$.

\change{
\subsection{Practical Usage}
We discuss the system level design and the use case for \systemname in a practical deployment. 

\subsubsection{Operation flow}
In the practical deployment, \systemname can use an ultra-low power FPGA device~\cite{lpfpga} to control the RF switch. Once the surface is attached to a specific object, the FPGA needs to be re-programmed at the same time so that it can transmit the corresponding object identification information. During the operation, the FPGA outputs the programmed control signal to make the \systemname switch between the constructive and destructive states. On the self-driving car side, the radar transmits the chirp signal and measure the signal reflected back to the radar receiver. The self-driving radar reconstruct the detection map for every chirp it receives. By comparing the detection map across chirps, the radar system can decode the modulated on-off-keying reflection and get the the object identification information.

\subsubsection{Concurrent detection}
In the real deployment, multiple vehicles might sends out chirp signal at the same time, which might cause interference in the receiving signal among different self-driving car radar. \systemname is inherently resistance to the interference caused by concurrent detection. Since \systemname beamforms the signal back to its incident direction, self-driving car radars at different directions can safely receive the reflection of its own signal. 
}
\subsubsection{Use cases}

\change{\systemname is able to enhance the radar detection robustness for self-driving systems.  We can envision three practical use cases for \systemname in the real world. 
\begin{itemize}
    \item \textbf{Augment existing radar detection.} \systemname can enlarge the RCS of an object so that the self-driving car radar can detect the object reliably. Just like fluorescent clothing can increase the visibility of human being, in the real world, \systemname can be worn by pedestrians or cyclists to reduce the change of being hit by self-driving vehicles. It can also be attached to the surface of the vehicles so that they can be detected easily. 
    \item \textbf{Detect traffic lines} 
    In addition, \systemname can also be attached to the traffic lines so that the self-driving car radar is able to detect traffic line so that they can keep the right lane even during heavy rains when traffic lines become hard to see. 
    \item \textbf{Detect traffic sign}
    \systemname can also help self-driving radar  read the detail content of  traffic sign. Specifically, \systemname can transmit the type of traffic sign as well as the content of the traffic sign to the self-driving car radar (e.g. whether it is a stop sign or what is the speed limit). 
\end{itemize}
}

\begin{figure*}[t]
    \centering
    \begin{subfigure}[b]{0.33\linewidth}
        \centering
        \includegraphics[width=\textwidth]{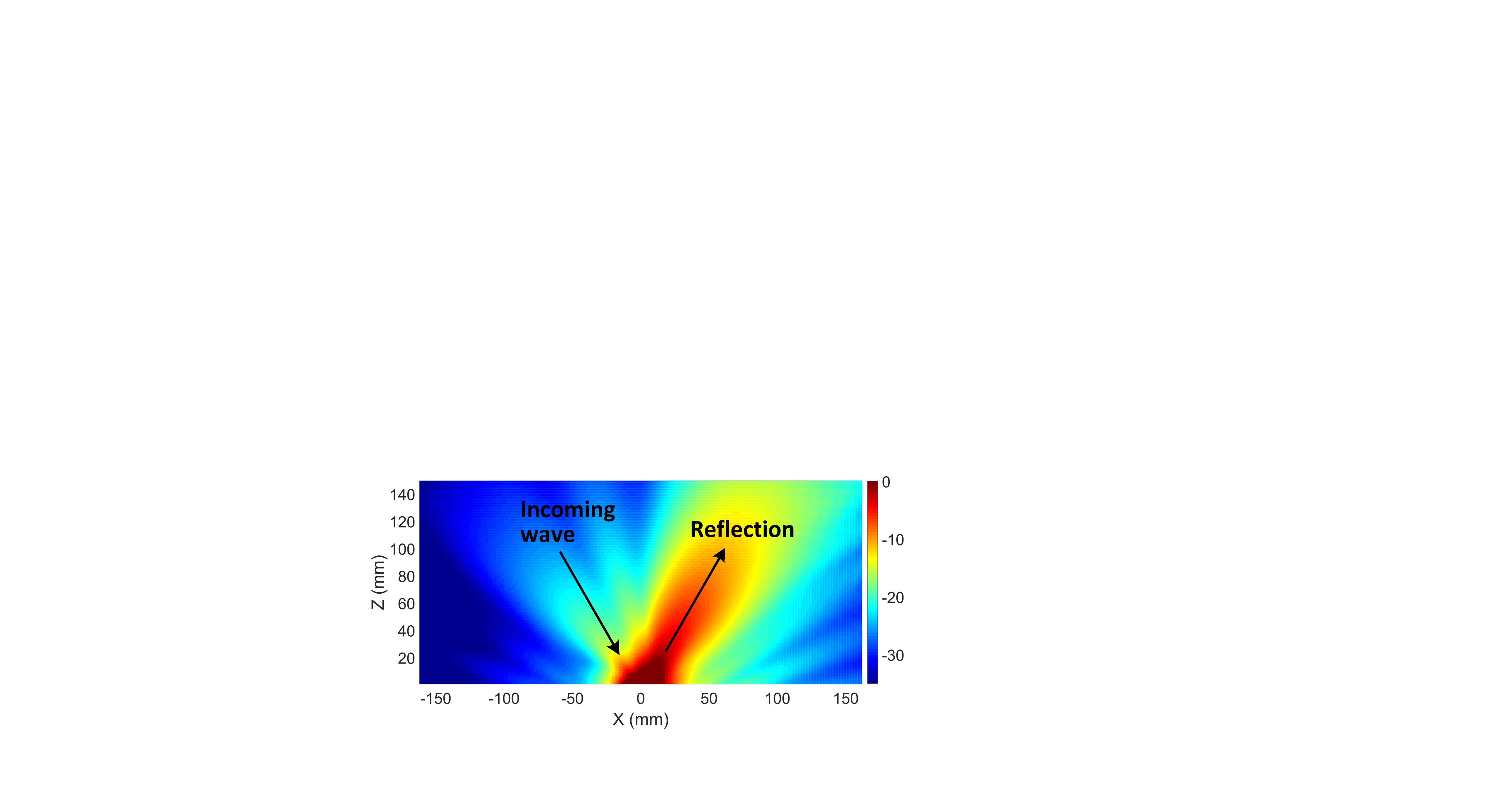}
        \caption{Metal surface baseline.}
        \label{fig:metal-baseline}
    \end{subfigure}
    \hfill
    \begin{subfigure}[b]{0.33\linewidth}
        \centering
        \includegraphics[width=\textwidth]{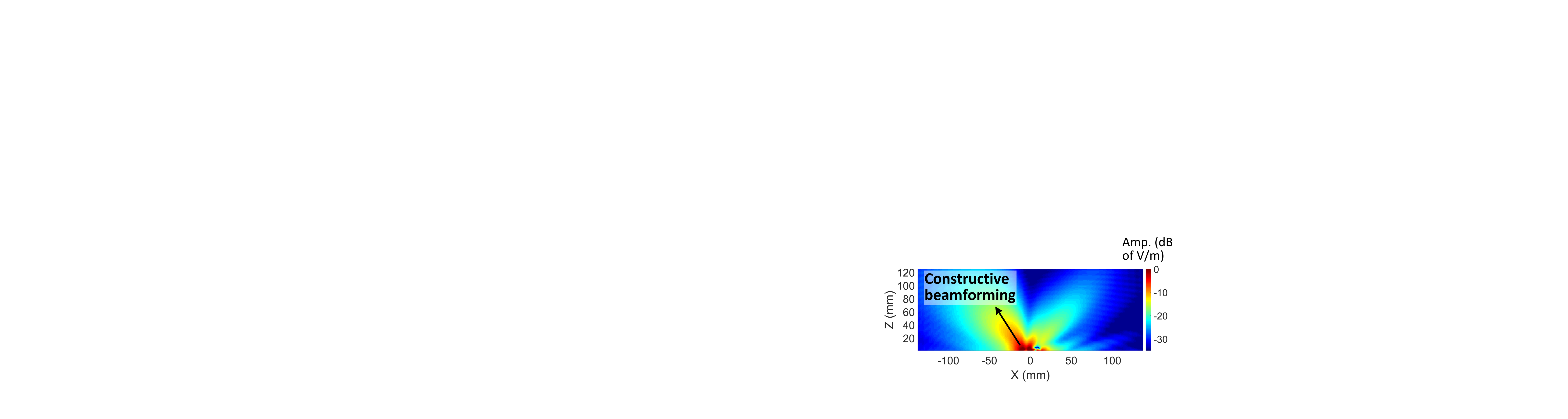}
        \caption{Constructive beamforming.}
        \label{fig:constructive}
    \end{subfigure}
    \hfill
    \begin{subfigure}[b]{0.33\linewidth}
        \centering
        \includegraphics[width=\textwidth]{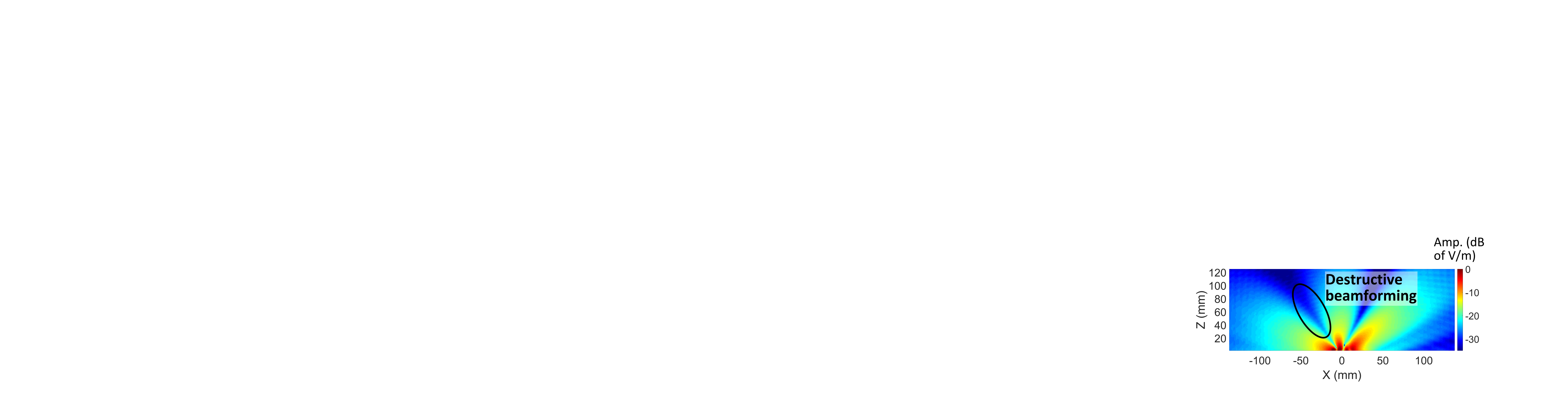}
        \caption{Destructive beamforming.}
        \label{fig:destructive}
    \end{subfigure}
    \caption{Simulated electric-field patterns for a four-element \systemname design.}
\label{fig:efield-simulation}
\end{figure*}




\begin{figure}[t]
\includegraphics[width=0.9\linewidth]{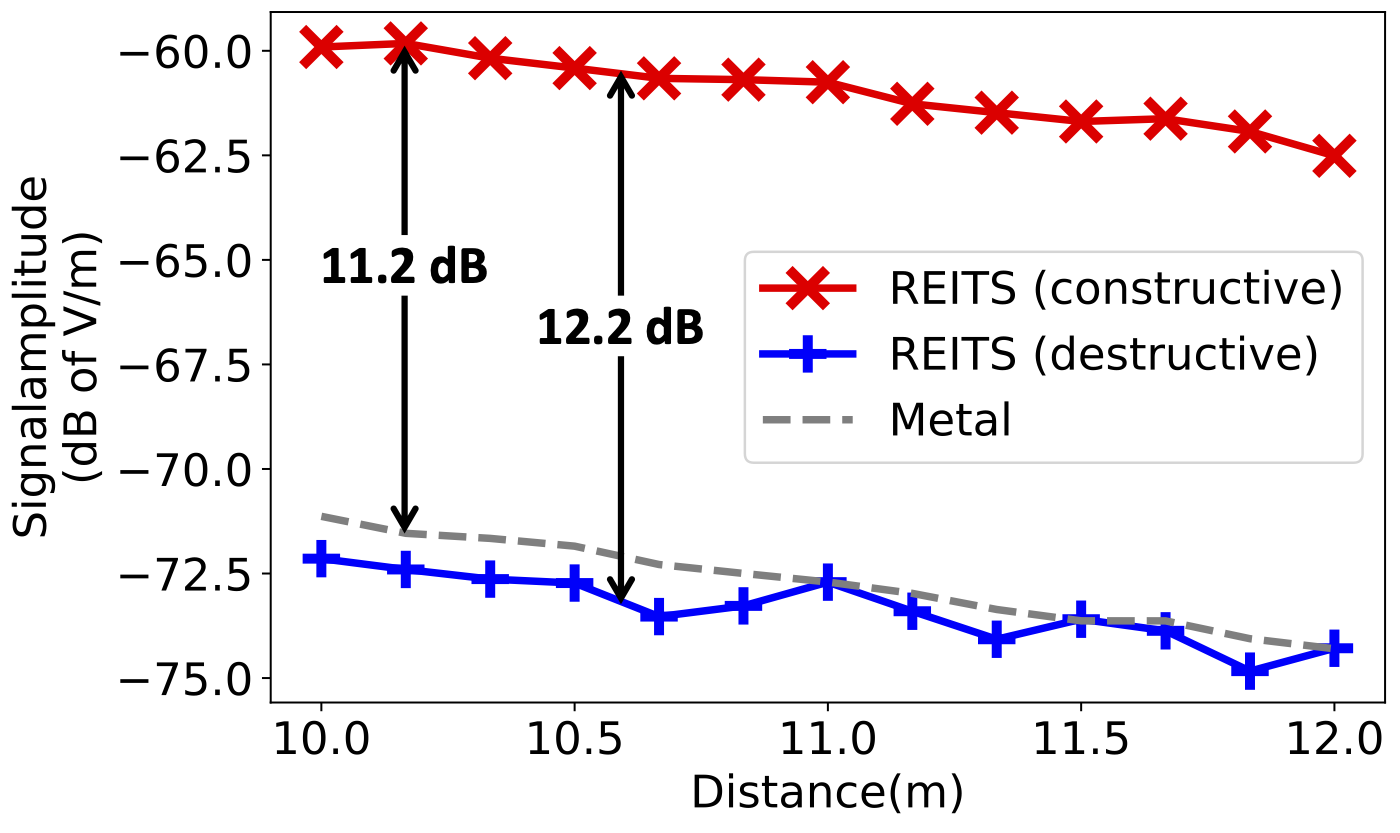}
\caption{Amplitude of the electric field in \\ the incoming direction ($\theta=30^o$) vs. distance, \\ when the incident amplitude is $1 V/m$.}
\label{fig:amplitude-distance}
\end{figure}

\section{Preliminary results}

\change{
We evaluate the design of \systemname with SIMULIA CST, a 3D electromagnetic simulator. In the simulation setup, 24 GHz plane wave excitation and open boundary condition are used to emulate a practical scenario, where the \systemname device reflects the signal from a radar in far field. The \systemname consists of a 0.338 mm-thick Rogers R4350B substrate, a $\lambda/2$-spaced linear array of metal patches (4.1 mm$\times$ 3.12 mm) as antennas, and a ground plane. The patch antennas are designed with $80\%$ radiation efficiency and $50 \Omega$ input impedance (realized by a 1.15 mm-long inset feed). Each centrosymmetric antenna pair is connected by a 0.67 mm-wide transmission line with a characteristic impedance of $50 \Omega$, for impedance match with the antennas. Controllable phase delay is achieved by assigning proper lengths to the transmission lines, for switching between the constructive and destructive settings.

We characterize the performance gain of \systemname against a blank metal baseline on 24-GHz self-driving radar frequency band. In addition, we examine the controllability of \systemname by comparing the signal strengths under constructive and destructive settings of the blind beamformer.  }

Figure~\ref{fig:efield-simulation} shows simulated electric-field patterns for \systemname. The three figures correspond to the cases where the incoming EM wave ($\theta=30^o$) is reflected by (1) a blank metallic rectangle; (2) a linear four-element beamformer in the constructive setting; (3) a linear four-element beamformer in the destructive setting. These devices share the same dimension (2.5 cm $\times$ 0.9 cm). As illustrated in Figure~\ref{fig:efield-simulation}(a), when hitting a metal object, the EM wave is reflected into the angle identical to that of the incoming wave ($\theta’=\theta$) but symmetric by the surface normal. When \systemname is set to the constructive state,
we can see a clear reflection peak at the incident direction, as shown in Figure~\ref{fig:efield-simulation}(b). 
When \systemname is set to the destructive state,
we can see a null at the incident direction, as shown in Figure~\ref{fig:efield-simulation}(c). 
Overall, when striking \systemname approximately $82\%$ of the EM wave power is absorbed by the four antennas and then re-radiated, forming either a constructive or a destructive beam at the incoming angle, depending on the switch settings.

Figure~\ref{fig:amplitude-distance} shows the amplitude of the electric field in the incoming direction (the data is extracted from a one-dimensional cut in Figure~\ref{fig:efield-simulation} with $\theta=30^o$, and then scaled to larger distances according to the $1/r$ scaling law for the electric field in far field \cite{EMfundamentals}). Compared to a blank metal surface, when \systemname is configured constructively, the return signal strength is enhanced by 11.2 dB. The gap indicates that \systemname can improve the detection distance of the target by $3.63\times$. The difference between constructive and destructive settings indicates a 12.2 dB modulation in signal amplitude. 

\begin{figure}[t]
\includegraphics[width=1\linewidth]{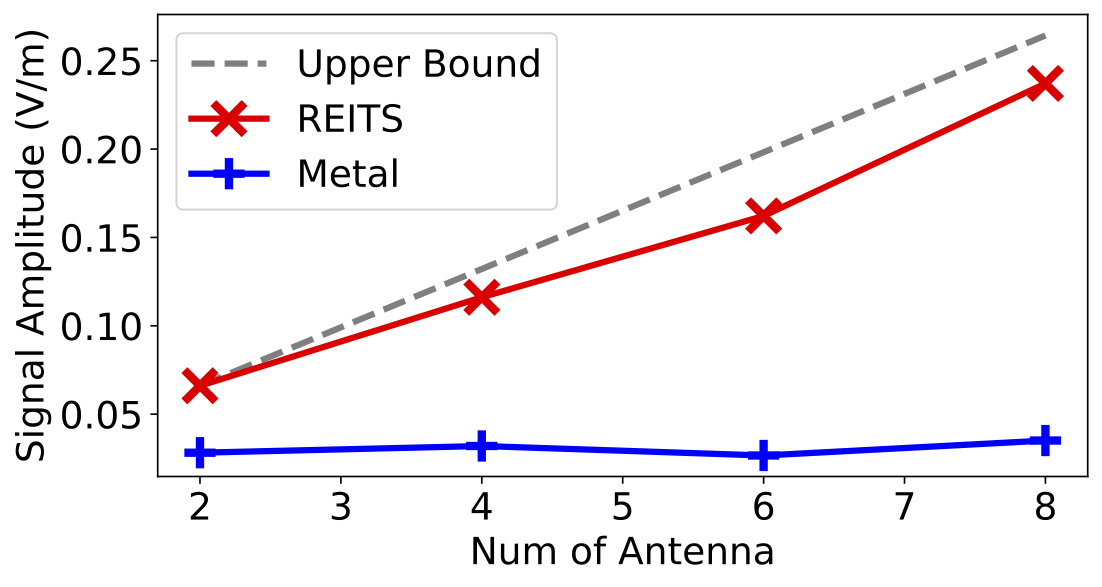}
\caption{The performance gain of \systemname \\ scales linearly with the number of antennas.}
\label{fig:amplitude-scale}
\end{figure}

We further investigate the scalability of our design, as shown in Figure~\ref{fig:amplitude-scale}. The grey dashed line is the upper bound, where the signal strength increases linearly with the number of antennas. We can see that \systemname can achieve close to linear scale-up as the number of antenna increases. In contrast, the signal reflection in the incident direction by a metal surface remains low. 

\balance

\section{Related Work}
\systemname is related to two lines of research. One is the research effort to 
improve the accuracy and robustness of self-driving sensing. Another is the endeavor for long-range passive communication. 

\subsection{Sensing for self-driving vehicles}
Self-driving vehicles depend on accurate and reliable sensing capability to detect obstacles on the road. Existing works have used vision-based methods~\cite{liang2018cirl,maqueda2018event} to extract the information from images captured by the camera. Vision-based methods can provide rich context information for self-driving vehicles. But it might fail due to adverse weather~\cite{tung2017raincouver} or cyber-physical attack~\cite{nassi2020phantom}. To deal with the failure of vision-based sensing, self-driving vehicles are also equipped with radars as a backup mechanism since it is more reliable in inclement weather. Many works also investigate the methods to improve the detection accuracy and robustness of self-driving vehicle radars~\cite{scheiner2020seeing,rohling2001waveform} with signal processing techniques. 

Different from existing work, which focuses on optimization from the view of self-driving vehicles, \systemname uses a different approach to improve the safety of self-driving vehicles from the infrastructure perspective. 
In addition, \systemname focuses on improving the signal itself. It is complementary to existing work, which applies various signal processing techniques to achieve accurate sensing.

\subsection{Passive communication design}
Research efforts also focus on the design and implementation of reliable passive tags for communication. One line of research is to extend the communication range of passive RFID tags. PushID~\cite{wang2019pushing} uses a distributed beamforming approach to extend the range of RFID communication. mmTag~\cite{mazaheri2020mill} uses a blind beamforming approach to increase the throughput of millimeter tag readers. The design of \systemname is also a type of passive hardware. Different from those, \systemname focuses on extending the range of sensing.

Another line of research pushes the range of passive tags to long-distance~\cite{peng2018plora,talla2017lora} by integrating backscatter techniques with LoRa. These techniques enable backscatter communication even if the signal strength is below the noise floor. But the tag should be deployed close to the excitation signal so that the backscattered signal can reach receivers at distance. 
\systemname targets on self-driving radar applications, where the distance from the tag to the signal source and the receiver are the same. At the same time, the latency of long-range backscatter is too high for the self-driving application, since long-range backscatter techniques have to aggregate dozens or hundreds of readings to boost the signal to noise ratio.

\section{Discussion}
In this paper, we have investigated the design of a programmable surface that enables robust sensing by self-driving vehicle radar. Our preliminary results show that the surface can extend the detection range of self-driving vehicles significantly. Nevertheless, in order to build a practical system based on the design the following topics require further study:

\parahead{Frequency trade-off}
Self-driving radar is licensed on two frequency bands by FCC: 24-GHz band and 76-GHz band. Currently, the design of \systemname is based on the 24-GHz frequency band. It is also meaningful to scale the design to 76-GHz frequency. At the higher frequency, the dimension of antenna is smaller. Therefore it can capture less energy with a single antenna. Besides, the high-frequency signal typically has a higher loss in the transmission line. In order to achieve the same performance gain, the surface at 76-GHz needs to have more antennas compared with the surface at 24-GHz. Therefore, we need a design that can deploy more antennas on a surface. 

\parahead{Multiple reflective surfaces}
The current design of \systemname focuses on improving the robustness of radar detection for a single \systemname surface. But in the real world, there might be multiple \systemname surfaces in the view of a self-driving vehicle. For two reflective surfaces that have enough separation in space, the radar system can detection both of than by looking at the difference of angle of arrival and time of flight. However, it is challenging to identify two object when they are too close to each other, when the object identification signal will interfere with each other. 
We need to continue to investigate how to distinguish those multiple surfaces even if they are located close to each other.

\parahead{Modulate Identification Information}
\systemname is able to transmit object identification information with on-off-keying. However, the modulation of the information also affects detection accuracy of the radar system. For example, if the frequency of constructive beamforming state is too sparse, the radar will receive a few chirps, which reduces the accuracy to compute the Doppler effect of the target object. 


\change{
\parahead{Doppler effect.}
The Doppler effect is introduced by the speed difference between the target and the self-driving vehicle. It might corrupt the object identification information. 
Since the radar chirp time is typical short (e.g. $0.5$ millisecond), \systemname can switch between constructive beamforming state and destructive beamforming state at a frequency where the Doppler effect is neglectable. For example, we can set the switch interval to one millisecond. The maximal acceleration of a typical car is around 3-6~$m^2/s$. The speed difference between two states is at most 0.003-0.006~$m/s$.
}

\section{Conclusion}
In this paper, we propose and verify the design of \systemname, a system to improve the robustness of detection of objects for autonomous vehicles by maximizing the RCS of the objects. 
\systemname adopts a multi-antenna design, which enables constructive blind beamforming to return an enhanced radar signal in the incident direction. \systemname can also let the existing radar system to sense identification information by switching between constructive  beamforming state and  destructive  beamforming state. 
Results of a preliminary evaluation show that \systemname enhances the return signal strength by 11.2~dB and extends the detection distance by $3.63\times$.

\begin{acks}
We thank the anonymous reviewers and our shepherd,
Haitham Hassanieh for their insightful comments. This work is  supported by the National Science Foundation under grant CNS-1617161, Semiconductor Research Corporation, Center for Brain-inspired Computing, and Princeton Program in Plasma Science and Technology (PPST).
Any opinions, findings, and conclusions or
recommendations expressed in this material are those of the
author(s) and do not necessarily reflect the views of the our funding agents.

\end{acks}

\bibliographystyle{acm}
\begin{raggedright}
\renewcommand{\bibfont}{\normalsize}
\bibliography{reference}
\end{raggedright}

\clearpage
\appendix

\end{document}